\documentstyle[12pt]{article}
\newcommand{\ra}{\rightarrow}
\def\dfrac#1#2{{\displaystyle {#1 \over #2}}}
\begin{document}
\pagestyle{empty}
\begin{flushright}
BU-HEP 94-29 \\
CERN-TH.7484/94 \\
ROME prep. 94/1050 \\
SHEP prep. 93/94-30
\end{flushright}
\centerline{\LARGE{\bf{Lattice Calculation of $D$- and $B$-meson}}}
\vskip 0.3cm
\centerline{\LARGE{\bf{Semileptonic Decays,}}}
\vskip 0.3cm
\centerline{\LARGE{\bf{using the Clover Action  at $\beta=6.0$ on APE }}}
\vskip 0.3cm
\centerline{\bf{C.R. ALLTON$^1$, M. CRISAFULLI$^1$, V. LUBICZ$^2$,
G. MARTINELLI$^{1,3}$, }}
\centerline{\bf{ F. RAPUANO$^{1,3}$, N. STELLA$^4$, A. VLADIKAS$^{5}$}}
\centerline{and}
\centerline{\bf{A. BARTOLONI$^1$, C. BATTISTA$^{1}$, S. CABASINO$^{1}$,
 N. CABIBBO$^{5}$,}}
\centerline{\bf{ E. PANIZZI$^1$, P.S. PAOLUCCI$^{1}$,
R. SARNO$^{1}$, G.M. TODESCO$^{1}$,}}
\centerline{\bf{ M. TORELLI$^{1}$,  P. VICINI$^{1}$}}
\centerline{The APE Collaboration}
\centerline{$^1$ Dip. di Fisica, Univ. di Roma \lq La Sapienza\rq and
INFN, Sezione di Roma,}
\centerline{P.le A. Moro, I-00185 Rome, Italy.}
\centerline{$^2$ Dept. of Physics, Boston University, Boston MA 02215,
USA.}
\centerline{$^3$ Theory Division, CERN, 1211 Geneva 23, Switzerland.}
\centerline{$^4$ Dept. of Physics, The University, Southampton SO9 5NH,
UK.}
\centerline{$^5$ Dip. di Fisica, Univ. di Roma \lq Tor Vergata\rq
and INFN, Sezione di Roma II,}
\centerline{Via della Ricerca Scientifica 1, I-00133 Rome, Italy.}
\date{}
\abstract{
\par\noindent
We present the results of a high statistics lattice calculation of hadronic
form factors relevant for $D-$ and $B-$meson semi-leptonic decays into
light pseudoscalar and vector mesons. The results have been obtained by
averaging over 170 gauge field configurations, generated in the quenched
approximation, at $\beta=6.0$, on a $18^3 \times 64$ lattice, using the
$O(a)$-improved SW-Clover action.
{}From the study of the matrix element $<K^-\vert J_\mu \vert D^0>$, we obtain
$f_+(0)=0.78\pm 0.08$ and from the matrix element $<\bar K^{* 0}\vert J_\mu
\vert D^+>$ we obtain $V(0)=1.08\pm 0.22$, $A_1(0)=0.67\pm 0.11$ and
$A_2(0)=0.49\pm 0.34$.  We also obtain the ratios $V(0)/A_1(0)=1.6\pm 0.3$ and
$A_2(0)/A_1(0)=0.7\pm 0.4$. Our predictions for the different form factors are
in good agreement with the experimental data, although, in the case of
$A_2(0)$, the errors are still  too large to draw any firm conclusion.
\par\noindent
With the help of the Heavy Quark Effective Theory (HQET) we have also
extrapolated the lattice results to  $B$-meson decays.
The form factors follow a behaviour compatible with the HQET predictions. Our
results are in agreement with a previous lattice calculation, performed at
 $\beta=6.4$, using the standard Wilson action.
}
\vskip 0.8cm
\begin{flushleft}
CERN-TH.7484/94 \\
November 1994
\end{flushleft}
\vfill\eject
\pagestyle{empty}\clearpage
\setcounter{page}{1}
\pagestyle{plain}
\newpage
\pagestyle{plain} \setcounter{page}{1}

\section{Introduction}
\label{sec:introduction}

There is increasing evidence that quantitative calculations of
weak decay amplitudes can be obtained by lattice $QCD$ simulations.
Over the last few years, semi-leptonic decays of heavy mesons have been studied
on the lattice  \cite{crisa}--\cite{gupta}.  Among
these, $D-$meson decays provide a good test of the lattice method,
since the relevant CKM matrix element is well constrained by unitarity in
the Standard Model, $V_{cs}\simeq 0.975$. The main advantage of the lattice
technique is that it is based on first principles only and it does not
contain free parameters besides the quark masses and the value of the
lattice spacing, both
of which are fixed by hadron spectroscopy. Moreover, statistical and systematic
errors in lattice simulations can be systematically reduced with
increasing computer resources.

In this
work, we present a high-statistics study of pseudoscalar-pseudoscalar and
pseu\-do\-sca\-lar-vector semi-leptonic form factors,  performed on
the 6.4 Gflops APE machine, by using the $O(a)$-improved Clover Action \cite
{sw,heatlie}, at $\beta=6.0$, corresponding to an inverse
lattice spacing $a^{-1} \sim 2$ GeV.
Our prediction for the  form factors, which govern the $D\rightarrow
K$ and $D\rightarrow K^{*}$  amplitudes are given in the abstract
and in table \ref{tab:formfac}. The central values are in remarkable
agreement with the experimental results. However, in spite of the large
statistics, our predictions still suffer, in some cases, from large
statistical errors. A possible explanation for the  errors' size
 is the use of a ``thinning''
procedure, that will be discussed in detail in sec. \ref{sec:systematic}.
This procedure was adopted because
 our computer memory is not sufficient to store the necessary quark
propagators.

Following the suggestion of ref. \cite{elc}, we have also tried
to extrapolate the form factors to $B\to \pi ,\rho $ decays, using the
scaling laws predicted by the Heavy Quark Effective Theory (HQET) \cite{iw}.
The final results have large uncertainties, because  the
statistical errors  amplify in the extrapolation. It is
reassuring  that the present results  are in good agreement
with those obtained, with the same method, in ref. \cite{elc}, by using the
standard fermion Wilson action,  at a smaller value of the lattice spacing,
corresponding to
 $\beta =6.4$. This gives us  confidence on the feasibility
of the extrapolation. It should be noticed however, that  the
form factors are obtained near $q^2 = q^2_{max}$, where  $q^2$ is the
momentum of the lepton pair.  In order to predict the form factors
at all $q^2$, on current lattices  and in the range of the
heavy quark masses explored so far,  we can only assume pole dominance (or
 any other simple $q^2$-dependence). This strongly biases the final results,
e.g. the determination
 of the
form factors at $q^2=0$. In this respect, the situation is not very different
from the quark model approach.
In order to improve the accuracy of the predictions,
it is necessary to be able to work with heavier quark masses and to increase
the
range of $q^2$. This can only be achieved  by going to larger lattices.
The approach followed here indicates that $B-$meson semi-leptonic decays will
be a fruitful area of lattice investigations in the near future.

\section{Computation Details}

\label{sec:details}

Semi-leptonic decays ($D\to K,K^{*}$; $B\to D,D^{*}$; $B\to \pi ,\rho $) are
described in terms of six independent dimensionless form factors, four
of which are important for  decay rates into light  leptons (see
e.g.  \cite{lms}). For each form factor, the relevant information can be
expressed in terms of its value  at $q^2=0$ and its
$q^2$ dependence. In this section several details of our calculation are given.

\begin{figure}[t]   
\begin{center} \setlength{\unitlength}{1truecm}
\begin{picture}(6.0,6.0)
\put(-6.0,-6.2){\special{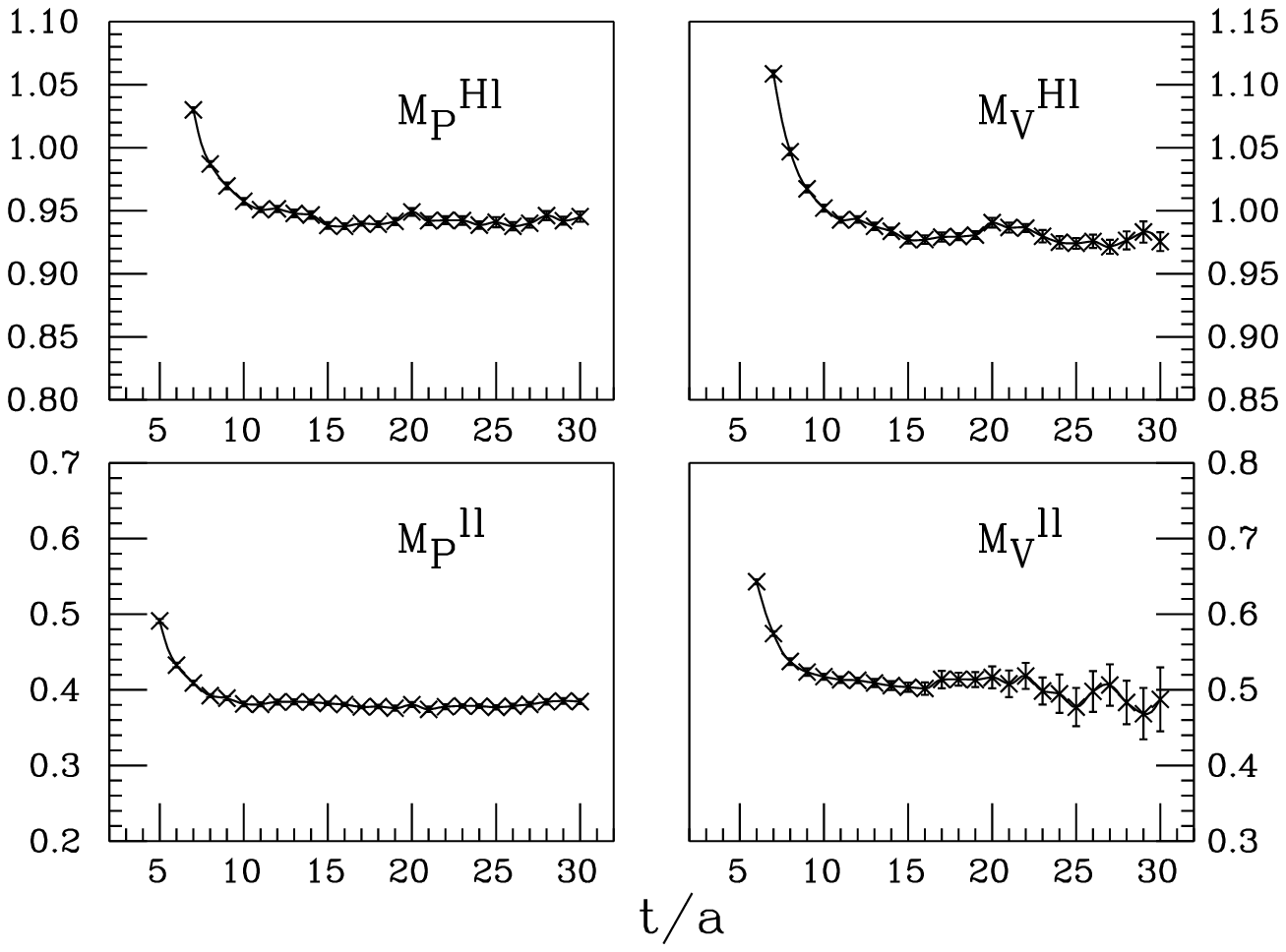}}
\end{picture} \end{center}
\vskip 2.6cm
\caption{\it{The effective mass, defined as $M(t)a=\ln(C(t)/C(t+a))$
as a function of $t/a$. $M_{P,V}^{Hl}$ and $M_{P,V}^{ll}$ refer to
the heavy-light
or the light-light pseudoscalar and vector mesons respectively.
$C(t)$ is the generic, zero  momentum two-point correlation function.}}
\label{fig:em}
\end{figure}

We have obtained our results, by averaging over 170 gauge field
configurations, generated at $\beta =6.0$, on a volume $18^3\times 64$. On
this set of configurations, we have computed the  quark propagators
by using the $O(a)$-improved SW-Clover action  \cite{sw,heatlie}. We have
considered 3 different values of the Wilson hopping parameter for the light
quarks, $K_l=0.1425$, $0.1432$ and $0.1440$, and 4 different values for the
heavy quarks, $K_H=0.1150$, $0.1200$, $0.1250$ and $0.1330$. The
values of $K$, corresponding to the  zero quark mass
and to the strange quark mass (obtained by fixing the pion and the kaon
pseudoscalar masses) are  $K_{cr}=0.14545(1)$ and $K_s=0.1435(2)$
respectively. The value of the charm quark mass, obtained from the $D-$meson
mass, corresponds to $K_{ch}=0.1219(17)$. The inverse lattice spacing,
obtained by using the mass of the $\rho -$meson to set the scale, is $%
a^{-1}=(1.95\pm 0.08)$ GeV.  All two-point functions
have been fitted in the interval $t/a=12-28$ and $15-28$ for light-light
and heavy-light mesons respectively. In fig. \ref{fig:em}, we give
examples of the effective mass, as a function of  time, in the
two cases.
The statistical errors
have been estimated with the jacknife method, by decimating 10
configurations from the total set of 170.
 Preliminary results obtained with a smaller set of
gauge field configurations  have been reported in ref. \cite{stella}.

To extract the form factors, we have computed two- and three-point
correlation functions and followed the procedure explained in ref. \cite{elc}.
The matrix elements have been computed using the pseudoscalar density
 as source for  $D$ and $K$ mesons and
the local vector current    for the $K^{*}$ meson\footnote{Throughout this
paper,
$K$, $K^*$ and $D$ are conventional names to denote the light pseudoscalar
and vector mesons and the heavy meson respectively.}.
 For
the weak current  we have used the lattice improved local vector and axial
vector currents, according to the definition
of refs. \cite{msv,mssv}.
 The corresponding renormalization constants, $Z_V$ and $Z_A$, have
been fixed to the values $Z_V=0.824$ and $Z_A=1.06$, as determined
non-perturbatively, using light quark correlation functions
\cite{mpsv,gribov}.
 This choice will be justified below.

All matrix elements have been computed by inserting the $D$-meson source at
a time distance $(t_D-t_{K,K^{*}})/a=28$. The position of the light meson is
fixed at the origin and we have varied the time position of the weak current
in the time interval $t_J/a=10-14$.

To study the $q^2$-dependence of the form factors, we have computed the
matrix elements in two different kinematical configurations: in the first case
we have taken the $D-$meson at rest, i.e. $\vec p_D=\vec 0$, and $\vec
p_{K,K^{*}}\equiv 2\pi /La\cdot (0,0,0)$, $(1,0,0)$, $(1,1,0)$, $(1,1,1)$
and $(2,0,0)$; in the second case we have chosen
 $\vec p_D\equiv 2\pi /La\cdot (1,0,0)$ and $%
\vec p_{K,K^{*}}\equiv 2\pi /La\cdot (0,0,0)$, $(1,0,0)$, $(-1,0,0)$, $%
(0,1,0)$. We have therefore nine independent momenta.
We have also computed other correlation functions,
 which are equivalent under the
cubic symmetry. All such equivalent cases  have been averaged together.

In order to extract the current matrix elements from the three-point
functions, we have used two different procedures, denoted by ``analytic'' and
``ratio'' methods, discussed in detail in ref. \cite{elc}. Within the
statistical errors, the above methods are expected to agree, up to $O(a)$
effects. In the present calculation, we find that the
two methods yield slightly different results, the differences varying
between $2\%$ and $8\%$.  We have taken into account these differences in
the evaluation of the final error (see below).

In order to obtain the form factors at $q^2=0$, for quark masses
corresponding to the physical $D$ ($B$) and $K$ ($\pi $) or $K^{*}$ ($\rho $%
) mesons,  we have extrapolated the form factors, both in masses and momenta.
The following procedure has been used:
\begin{itemize}
\item[1)] At fixed heavy quark mass and  meson momenta,
$\vec p_{K,K^{*}}$%
, the generic form factor $F$ ($F=f_{+}$, $A_1$, $A_2$ and $V$) has been
extrapolated linearly in the light quark mass, to values corresponding to the
strange ($D\rightarrow K,K^{*}$) or massless ($D\rightarrow \pi ,\rho $)
quarks.
\item[2)] $F$ has been extrapolated in the mass of the heavy meson,
 to the
$D-$ and $B-$meson masses, using the dependence expected in the HQET (see
eqs.(\ref{scala}) below). In order to evaluate the stability of these results
with
respect to a different extrapolation, we have also extrapolated $F$
in the mass of the heavy meson $M_P$ according to the ``na\"\i ve'' expression
$%
F=\alpha+\beta /M_P$. The differences between the two methods are
discussed later on.
\item[3)] In order to obtain
the form factors at $q^2=0$, we have only used
the points with  $\vec p_D=0$ and $\vec p_{K,K^{*}}=
2\pi /La\cdot (1,0,0)$, which have been extrapolated  by assuming
meson pole dominance, $F(q^2)=F(0)/(1-q^2/M_t^2)$.
This reduces the uncertainty of the extrapolation because,
in most of cases, the point with $\vec p_{D}=2\pi /La\cdot (0,0,0)$  and
$\vec p_{K,K^{*}}=2\pi /La\cdot (1,0,0)$ corresponds to the smallest $q^2$.
$M_t$ is the mass of the lightest meson exchanged in the t-channel. Thus,
the vector and axial meson masses have been used for the extrapolation of
$f^{+},V$ and $A_{1,}A_2 $ respectively. $M_t$ has been computed on the
lattice, over the same gauge field configurations, at the same heavy and light
quark masses used for the three-point functions. In order to obtain the
physical $D-$ and $B-$meson masses, we have fitted the vector-pseudoscalar
mass difference $\Delta M=M_{P^{*}}-M_P$ (and similarly for the axial and
scalar cases) as $\Delta M=A_M+B_M/M_P$. The results, extrapolated to the
strange and massless light quarks, are reported in table \ref{tab:masses}.
These values have been used to extrapolate the form factors at $q^2=0$.
In this extrapolation, the precise value of $M_t$ is relatively unimportant.
For example we have verified that, by using the vector meson mass in all
cases, the results change by only a few per cent.
\end{itemize}
\begin{table}
\centering
\begin{tabular} {|c|c|c||c|c|c|}
\hline
{\em $M_{D^*}$} & {\em $M_{D^{**}}\, 1^{++}$ } & {\em $M_{D^{**}}\,
0^{++}$ } & {\em $M_{{D_s}^*}$ } & {\em $M_{{D_s}^{**}}\, 1^{++}$ } &
{\em $M_{{D_s}^{**}}\, 0^{++}$ } \\ \hline
$1.95 \pm 0.01$ & $2.29 \pm 0.34$ & $2.03 \pm 0.19$ &
$2.05 \pm 0.01$ & $2.40 \pm 0.34$ & $2.13 \pm 0.19$ \\
\hline \hline
{\em $M_{B^*}$} & {\em $M_{B^{**}}\, 1^{++}$ } & {\em $M_{B^{**}}\,
0^{++}$ } & {\em $M_{{B_s}^*}$ } & {\em $M_{{B_s}^{**}}\, 1^{++}$ } &
{\em $M_{{B_s}^{**}}\, 0^{++}$ }
\\ \hline
$5.32 \pm 0.01$ & $5.99 \pm 0.62$ & $5.46 \pm 0.32$ &
$5.46 \pm 0.01$ & $6.13 \pm 0.62$ & $5.60 \pm 0.32$ \\
\hline
\end{tabular}
\caption{\it{Masses  (in GeV)  of the vector,
axial and scalar excitations for the $D$ and $B$ mesons
as determined from our simulation. The experimental
pseudoscalar masses $M_D$ and $M_B$ are used as an input.
These masses have been used to extrapolate the form factors to  $q^2=0$.}}
\label{tab:masses}
\end{table}

\begin{figure}[t]   
\begin{center} \setlength{\unitlength}{1truecm}
\begin{picture}(6.0,6.0)
\put(-6.0,-6.2){\special{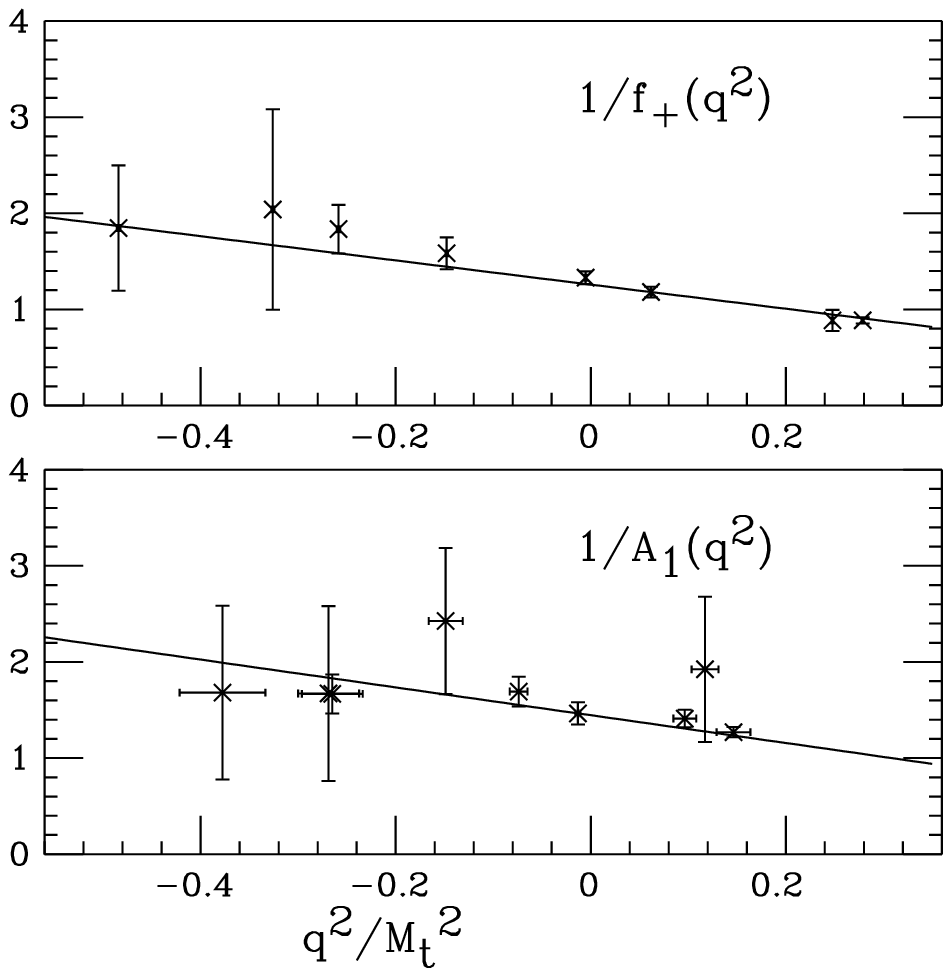}}
\end{picture} \end{center}
\vskip 2.6cm
\caption{\it{$1/f_+(q^2)$ and $1/A_1(q^2)$ as a function of the
dimensionless variable $q^2/M_t^2$. The heavy and light Wilson parameters
correspond to $K_H=0.1250$ and $K_l=0.1432$ respectively. The lines
represent the  pole dominance approximation.}}
\label{fig:poledom}
\end{figure}

The $q^2$-dependence of $1/f^+(q^2)$ and $1/A_1(q^2)$
 is compared with the meson
dominance predictions  in fig. \ref{fig:poledom}. The values of the inverse
form
factors are given as a function of the dimensionless variable $q^2/M_t^2$,
for the values of the heavy and light hopping parameter, $K_H=0.1250$ and
$K_l=0.1432$. The lines in the figures represent the pole dominance
expectations, with the pole masses computed on the lattice. In order
to reduce the error due to the extrapolation in $q^2$, the values of
the form factors at zero momentum transfer have been obtained, by fitting only
the  point closest to $q^2=0$, as explained above.  The other points
give,  however, important information on the $q^2$-dependence of the
form factors.
As shown in the figure, the $q^2$ behaviour
of the two form factors is  compatible with  pole dominance
predictions. Similar conclusions can be reached also for the form factors $%
V(q^2) $ and $A_2(q^2)$. Notice that
  the axial form factors determined by using $QCD$ sum
rules  \cite{bbd,ball} do to follow the pole
dominance behaviour.

\section{Main Sources of  Errors}
\label{sec:systematic}

In this section, we briefly describe the main sources of  error
which are present in our calculation, besides the quenched approximation.

\subsection{The ``Thinning'' Procedure}

``Thinning'' means that, when  computing the correlation functions,
 one  uses only one point out of $N_{th}$, in each spatial
direction. In our case $N_{th}=3$. Thus, for example,
the Fourier transform of the two-point correlation function is defined
 as
\begin{equation}
C(t,\vec{p}) = {\cal N_T}
 \sum_{\vec{x}_T} e^{-i\vec{p}\cdot\vec{x}_T} \, C(t,\vec{x}_T)
\end{equation}
where $x_T = 0,3,6, ...$ in each  direction and ${\cal N_T}$ is a suitable
normalization factor. This procedure is
necessary when, as in our case, the computer memory is not sufficient to
store the full quark propagators. There is a systematic error introduced by
thinning, because we cannot eliminate high momentum components in the
correlation functions. For $N_{th}=3$, it is possible to show that, for each
spatial direction, two higher momentum components ($p_1$ and $p_2$), besides
the smallest one ($p_0$), give a contribution to the correlation function.
$p_1$ and $p_2$ are related to $p_0$ by the simple relation:
\begin{equation}
\label{thin}p_m=p_0+\frac{2\pi }{La}\cdot \left( \frac L3\right) \cdot
m\qquad ,\qquad m=1,2
\end{equation}
The systematic error introduced by thinning is expected to be negligible at
large time distances, since the contribution of the unwanted higher energy
states  is exponentially suppressed in time. On the other
hand, the resulting signal may be  noisier,  because we use a small
sample of the lattice points.

In the case of the two-point correlation functions, we have been able to
directly compare ``thinned'' and ``non-thinned'' correlation functions,
computed on the same set of gauge configurations, and for momenta
$\vec p=2\pi /La\cdot
(0,0,0)$ and $(1,0,0)$. For the pseudoscalar and
vector meson correlations,  no observable statistical or systematic
effect has been detected. However, the statistical noise  introduced by
thinning  may  be larger, when we compute three-point correlation
functions, simply because, in the latter case,
 we have  to thin twice. Thus, the thinning procedure could be
responsible for  the quite
large statistical errors which have been found, in spite of the high
statistical
sample used in this calculation. Another reason could be the small spatial
volume of our lattice.

\subsection{$O(g^2a)$ and $O(a^2)$ Effects}

Using the Clover action, discretization errors  are
of  $O(g^2a)$ and $O(a^2)$.
In the case of light quarks,  these effects
have been shown to be much smaller than in the case of the
standard Wilson  action  \cite{msv}.  In the charm quark mass region,
at $\beta \sim 6.0$, however, discretization errors may still be important
also in the Clover case. \par
An estimate of the lattice artefacts can be obtained by comparing values
of lattice  renormalization constants, computed
 from different matrix elements, in a non-perturbative way
 \cite{msv}. In ref. \cite{newwi},  the
renormalization constants of the local vector and axial vector currents,
$V_\mu^L=\bar \psi_1 \gamma_\mu \psi_2$  and
$A_\mu^L=\bar \psi_1 \gamma_\mu \gamma_5 \psi_2$, have been determined,
at $\beta=6.0$,
as a function of  the quark masses, $m_1$ and $m_2$,
using the Ward identity method
\cite{msv,mpsv}. A variation  of about
$10-15 \%$,  on  the values of $Z_V$ and $Z_A$, has been observed,
for heavy quark masses between $m_H a =0.3$ and $0.9$, which is the
range used in the present study. The renormalization of
the quark fields, proposed in ref. \cite{kronfeld} in order to reduce
$O(m_H a)$ effects, enters, in our case, only at $O(m_H^2 a^2)$,
since we are using an improved action and improved operators
\cite{sw,heatlie}.  Since the effects that we have observed are linear
in $m_H a$, the KLM
correcting factor \cite{kronfeld} is of no help in our case. We
interpret the residual mass  dependence as an effect of $O(\alpha_s a)$.
Similar results
have been observed by the UKQCD Collaboration, which works with a Clover
action, at $\beta=6.2$ \cite{ukqcd}. For the above reasons, we have used,
through this paper, the non-perturbative values of $Z_V$ and $Z_A$, as
determined in refs. \cite{mpsv,gribov}, for light quark masses. A $10-15 \%$
of systematic error is then certainly present in our final results,
due to  residual $O(\alpha_s a)$ effects.

\subsection{Extrapolations of the Form Factors}

As explained in the previous section, the form factors, computed in a
range of heavy quark masses around the charm, must be extrapolated in
$1/M_P$ and $q^2$, in order to
obtain predictions for the physical form factors at zero momentum transfer.
There is also
an  extrapolation in the light quark mass, but this
 is quite smooth and unlikely to
be a source of an important uncertainty, within the present statistical
accuracy.
The extrapolation of the form factors in $1/M_P$ is more
delicate. There are
arguments, based on HQET, which allow the expansion of the form factors in
inverse powers of the heavy meson mass $M_P$. On the basis of HQET, up to $%
O(1/M_P^2)$ and up to logarithmic corrections, one expects the following
behaviour for the relevant form factors \cite{iw}:
\begin{equation}
\label{scala}
\begin{array}{ll}
\dfrac{f^{+}}{\sqrt{M_P}}=\gamma _{+}\times \left( 1+\dfrac{\delta _{+}}{M_P}%
\right) \qquad & \dfrac V{
\sqrt{M_P}}=\gamma _V\times \left( 1+\dfrac{\delta _V}{M_P}\right) \\  &  \\
A_1\sqrt{M_P}=\gamma _1\times \left( 1+\dfrac{\delta _1}{M_P}\right) \qquad
& \dfrac{A_2}{\sqrt{M_P}}=\gamma _2\times \left( 1+\dfrac{\delta _2}{M_P}%
\right)
\end{array}
\end{equation}
The expansions given in eqs. (\ref{scala}) are valid, in the limit of large
heavy quark mass, at fixed momentum $\vec p$ of the light meson (in the
frame where the heavy meson is at rest) when $|\vec p|\ll M_P$. The above
conditions are always satisfied for $q_{max}^2$, when the initial and final
mesons are both at rest. In our simulation, they are also satisfied for
the points corresponding to $\vec p=2\pi /La\cdot (1,0,0)$ (these have been
used in order to obtain all our final predictions).

In sec. \ref{sec:results},  we show that the dependence of the form factors on
the heavy meson mass, $M_P$, is compatible with the HQET predictions.
In order to evaluate the stability of the results with respect to a
different extrapolation, we have also  used a ``na\"\i ve'' scaling law of the
form:  \begin{equation}
\label{naive}f^{+},\,V,\,A_1,\,A_2=\alpha \times \left( 1+\dfrac \beta
{M_P}\right)
\end{equation}
For $D-$mesons, we find that the differences between the different
extrapolations, eq.(\ref{scala}) or (\ref{naive}), are quite small ($\le
2\%$) and completely negligible with respect to the statistical errors. The
results for $B-$meson decays  will be reported in sec. \ref{sec:results}.
In this case the differences are larger, of the order of $10-20\%$,
but still small with respect to the statistical errors, which amplify in the
extrapolation. Thus, we are not able to distinguish between the two
different behaviours.

We now turn the discussion to the $q^2$ extrapolation.
In the range of heavy quark masses considered in this study, the
$q^2$-dependence of the form factors is
 compatible with  pole dominance predictions. This range
corresponds approximately
to a heavy meson mass of the order of the $D-$meson mass and we also have
points
at $q^2\sim 0$. Therefore, as far as the $D$ decays are
concerned, the extrapolation to $q^2=0$ does not represent an important
source of theoretical uncertainty.
The extrapolation in $1/M_P$ to the $B$-meson mass
results in   momentum transfers
close to the maximum one. In order
to then predict the form factors at small $q^2$,
we have been forced to assume pole dominance. Since the range
 of the extrapolation
is very large\footnote{While the typical $\sqrt{q^2}$ on our lattice is at most
1.6 GeV, the extrapolation in the heavy quark mass will bring us to $\sqrt{q^2%
}=4.5$ GeV ($q^2=M_B^2-2M_B\sqrt{M_\pi ^2+(2\,\pi /18\,a)^2}+M_\pi ^2$), and
similarly for $B\rightarrow \rho $.},
this strongly biases our final results, e.g. the determination of the
form factors at $q^2=0$. In particular, we have used pole dominance
also in those cases, where it is in contradiction
with the results of $QCD$ sum rules calculations \cite{bbd,ball}.
 In order to improve the situation,
it is necessary to be able to work with heavier quark masses and to increase
the
range of $q^2$. This can only be achieved  by going to larger lattices
(at larger values of $\beta$).

\subsection{The ``Analytic'' and ``Ratio'' Methods}

In order to extract the weak current
matrix elements from the three-point functions, we used two different
procedures, denoted by ``analytic'' and ``ratio'' methods. The two
procedures are discussed in detail in ref. \cite{elc}.  The
``ratio'' methods means that, at each fixed time distance, we divide the
three-point correlation function by the two relevant two-point functions of
the $D$ and $K(K^{*})$ meson (with corresponding momentum) in order to cancel
 the
exponential time-dependence of the three-point correlation function.
 The ``analytic'' method differs from the
previous one because, instead of dividing by the two-point correlation
functions, computed at  different momenta, we divide by the
corresponding analytical expressions, with the source matrix elements and the
meson energies (computed from the meson masses, see eqs.(\ref{eq:enl})
and (\ref{eq:enc}) below) taken from the
fit of the two-point functions, at zero momentum.

When both the initial and final mesons are at rest these two methods are
practically equivalent and lead to almost identical results. However, when
the meson momenta are different from zero, within the statistical fluctuations,
the two methods are expected to agree only up to $O(a)$ effects.
\par  At $\beta=6.2$ and with a lattice volume
$24^3 \times 48$, it was found that discretization
errors seem to be reduced, by using, for
the two-point correlation
functions,   the  ``lattice" dispersion relation of a free boson
\begin{equation} \bar C(t,\vec p) = \frac{Z}{2 {\rm sinh} E} e^{-E t},
\label{eq:emr} \end{equation}
where
\begin{equation} E=\frac{2}{a} {\rm arc sinh}
\left( \sqrt{{\rm sinh}^2 ( \frac{m a}{2}) +\sum_{i=1,3}{\rm sin}^2(\frac
{p_i a}{2})}
\right) \label{eq:enl} \end{equation}
 and $\vec p$ is the momentum of the meson \cite{juan}. The same is true in our
case. We have verified this point, by studying
 the ratio  $R(t)= C(t,\vec p)/\bar C(t,\vec p )$.
 $C(t, \vec p)$ is the two-point correlation function of a meson with
momentum $\vec p$, as computed in our  simulation.   $\bar C(t, \vec p)$
is the expression in eq. (\ref{eq:emr}), with $Z$ and $m$ taken
from the fit of  the zero momentum correlation to $\bar C(t, \vec p=\vec 0)$.
At large time distances, we find that $R(t)$ is  close to one
(typically $1.05 \pm 0.01$). This compares favourably to the
value  $1.10 \pm 0.02$ in the case where,
instead of  eq. (\ref{eq:emr}), we use  the standard expression
\begin{equation} \hat C(t,\vec p) = \frac{Z}{2 E} e^{-E t},
 \end{equation}
with
\begin{equation} E=\sqrt{m^2 + \vert \vec p\vert ^2}.\label{eq:enc}
\end{equation} For this reason, in all our analysis, we have fitted the two
point functions to $\bar C(t,\vec p )$, as defined in eq. (\ref{eq:emr}).
 \par
Using $\bar C(t, \vec p)$, the ``ratio" and ``analytic"
methods yield slightly different results.
 In order to exhibit the differences, we
give in table \ref{tab:anaratio} the corresponding sets of values for
the form factors $f_{+}(0)$, $V(0)$, $A_1(0)$ and $A_2(0)$, as well as for
the ratios $V/A_1$ and $A_2/A_1$ at zero momentum transfer, for the $%
D\rightarrow K$ and $D\rightarrow K^{*}$ decays.
\begin{table}
\centering
\begin{tabular}{|c|c|c|c|c|}
\hline
& $f_+(0)$ & $V(0)$ & $A_1(0)$ \\ \hline
analytic & $0.81 \pm 0.07$ & $1.07 \pm 0.21$
         & $0.66 \pm 0.11$ \\ \hline
ratio    & $0.75 \pm 0.06$ & $1.09 \pm 0.22$
         & $0.67 \pm 0.11$ \\ \hline \hline
&$ A_2(0)$ & $V(0)/A_1(0)$ & $A_2(0)/A_1(0)$ \\ \hline
analytic & $0.44 \pm 0.34$ & $1.63 \pm 0.29$
         & $0.67 \pm 0.43$ \\ \hline
ratio    & $0.52 \pm 0.29$ & $1.62 \pm 0.29$
         & $0.79 \pm 0.35$ \\ \hline
\end{tabular}
\caption{\it{Semi-leptonic form factors  at zero momentum transfer for
$D \ra K$ and $D\ra K^*$ decays. The results have been obtained by
using the method called ``analytic" or ``ratio" to extract the form factors.}}
\label{tab:anaratio}
\end{table}
{}From the entries in the table one can see that the differences are larger for
the decay in the pseudoscalar channel and they
are always between $2\%$ and $\sim
8\%$\footnote{
For the form factor $A_2$, the difference is of the order of $15\%$. In
this case, however, the statistical errors are so large that
 the possibility of drawing any conclusion is prevented.}.
 In our final results,
the differences between the two methods have been added in quadrature with the
statistical errors.
Had we used the standard expression $\hat C(t, \vec p)$, we would had found
larger differences between the ``ratio" and ``analytic" methods, which
in the worst case have been evaluated to be $11 \%$, see also ref.
\cite{elc}.

\section{Physics Results}
\label{sec:results}

\subsection{$D-$Meson Decays}

Our best estimates for the
form factors and partial widths are those given in the abstract and in
tables \ref{tab:formfac} and \ref{tab:larghezze}. In these tables, we present
our results, together with other calculations and experimental determinations
of the form factors.%
\begin{table}
\centering
\begin{tabular}{|c|c||c|c|c|}
\hline
\multicolumn{2}{|c||}{\bf Reference} & {\bf f$_+$(0)} & {\bf V(0)} & {\bf
A$_1$(0)} \\
\hline \hline
EXP & Average \cite{with} &
$0.77 \pm 0.04$ & $1.16 \pm 0.16$ & $0.61 \pm 0.05$ \\ \hline  
LAT & {\bf This work} &
$0.78 \pm 0.08$ & $1.08 \pm 0.22$ & $0.67 \pm 0.11$ \\
& LMMS \cite{lmms} &
$0.63 \pm 0.08$ & $0.86 \pm 0.10$ & $0.53 \pm 0.03$ \\
& BKS \cite{bks} &
$0.90 \pm 0.08 \pm 0.21$ & $1.43 \pm 0.45 \pm 0.49$ & $0.83 \pm 0.14 \pm
0.28$ \\
& ELC \cite{elc} &
$0.60 \pm 0.15 \pm 0.07 $ & $0.86 \pm 0.24$ & $0.64 \pm 0.16$ \\
& UKQCD \cite{ukqcd} &
$0.67^{+0.07}_{-0.08}$ & $1.01^{+0.30}_{-0.13}$ & $0.70^{+0.07}_{-0.10}$ \\
& LANL \cite{gupta} &
$0.73 \pm 0.05$ & $1.27 \pm 0.08$ & $0.66 \pm 0.03$ \\ \hline
SR & BBD \cite{bbd} &
$0.60^{+0.15}_{-0.10}$ & $1.10 \pm 0.25$ & $0.50 \pm 0.15$ \\ \hline
QM & WSB \cite{wsb} &
$0.76$ & $1.23$ & $0.88$ \\
& ISGW \cite{isgw} &
$0.8$ & $1.1$ & $0.8$ \\
& GS \cite{gs} &
$0.69$ & $1.5$ & $0.73$  \\ \hline \hline
\multicolumn{2}{|c||}{\bf Reference} & {\bf A$_2$(0)} & {\bf
V(0)/A$_1$(0)} & {\bf A$_2$(0)/A$_1$(0)} \\ \hline \hline
EXP & Average \cite{with} &
$0.45 \pm 0.09$ & $1.90 \pm 0.25$ & $0.74 \pm 0.15$ \\ \hline  
LAT & {\bf This work} &
$0.49 \pm 0.34$ & $1.6 \pm 0.3$ & $0.7 \pm 0.4$ \\
& LMMS \cite{lmms} &
$0.19 \pm 0.21$ & $1.6 \pm 0.2$ & $0.4 \pm 0.4$ \\
& BKS \cite{bks} &
$0.59 \pm 0.14\pm0.24$ & $1.99 \pm 0.22 \pm 0.33$ & $0.7 \pm 0.16 \pm 0.17$ \\
& ELC \cite{elc} &
$0.40 \pm 0.28 \pm 0.04$ & $1.3 \pm 0.2$ & $0.6 \pm 0.3 $ \\
& UKQCD \cite{ukqcd} &
$0.66^{+0.10}_{-0.15}$ & $1.4^{+0.5}_{-0.2}$ & $0.9 \pm 0.2$ \\
& LANL \cite{gupta} &
$0.44 \pm 0.16$ & $1.83 \pm 0.09$ & $0.74 \pm 0.19$ \\ \hline
SR & BBD \cite{bbd} &
$0.60 \pm 0.15$ & $2.2 \pm 0.2$ & $1.2 \pm 0.2$ \\ \hline
QM & WSB \cite{wsb} &
$1.15$ & $1.4$ & $1.3$ \\
& ISGW \cite{isgw} &
$0.8$ & $1.4$ & $1.0$ \\
& GS \cite{gs} &
$0.55$ & $2.0$ & $0.8$ \\ \hline
\end{tabular}
\caption{\it{Semi-leptonic form factors for $D \ra K$ and $D\ra K^*$ decays.
The label EXP refers to the experimental results and the labels LAT, QM and
SR correspond to lattice, quark model and sum rules calculations
respectively.}}
\label{tab:formfac}
\end{table}
\begin{table}
\centering
\begin{tabular}{|c|c||c|c|c|}
\hline
\multicolumn{2}{|c||}{Reference} & $\Gamma(D\ra K)$ & $\Gamma(D\ra K^*)$ &
$\dfrac{\Gamma_L(D\ra K^*)}{\Gamma_T(D\ra K^*)}$ \\ \hline \hline
EXP & Average \cite{with,stone} &
$9.0 \pm 0.5$ & $5.1 \pm 0.5$ & $1.15 \pm 0.17$ \\ \hline
LAT & {\bf This work} &
$9.1 \pm 2.0$ & $6.9 \pm 1.8$ & $1.2 \pm 0.3$ \\
& LMMS \cite{lmms} &
$5.8 \pm 1.5$ & $5.0 \pm 0.9$ & $1.51 \pm 0.27$ \\
& ELC \cite{elc} &
$5.4\pm 3.0\pm 1.4$ &  $6.4\pm 2.8 $ & $1.4 \pm 0.3 $ \\
& UKQCD \cite{ukqcd} &
$7.0\pm 1.6\pm 0.4$ & $6.0^{+0.8}_{-1.6}$ & $1.06\pm 0.16\pm 0.02$ \\
 \hline  SR & BBD \cite{bbd} &
$6.4 \pm 1.4$ & $3.8 \pm 1.5$ & $0.86 \pm 0.06$ \\ \hline
QM & WSB \cite{wsb} &
$8.8$ & $9.7$ & $0.92$ \\
& ISGW \cite{isgw} &
$8.5$ & $9.2$ & $1.09$ \\
& GS \cite{gs} &
$7.1$ & $-$ & $-$ \\  \hline \hline
\multicolumn{2}{|c||}{Reference} & $\Gamma(D^0\ra\pi^-)$ & $\Gamma(D^0\ra
\rho^-)$ &
$\Gamma(D_s\ra\phi)$ \\
\hline \hline
EXP & Average \cite{pdg} &
$0.9^{+0.3}_{-0.2}$ & -- & $4.0 \pm 0.6$ \\ \hline
LAT & {\bf This work} &
$0.8 \pm 0.2$ & $0.6 \pm 0.2$ & $6.4 \pm 1.1$ \\
& LMMS \cite{lmms} &
$0.5 \pm 0.2$ & $0.40 \pm 0.09$ & $4.4 \pm 0.6$ \\
& ELC \cite{elc} &
$0.5\pm 0.3\pm 0.1 $ & $0.6\pm 0.3 \pm 0.1$ & -- \\
& UKQCD \cite{ukqcd} &
$0.52\pm 0.18\pm 0.04$ & $0.43 \pm 0.11$ & -- \\  \hline
SR &  Ball \cite{ball} &
$0.39 \pm 0.08$ & $0.12 \pm 0.03$ & -- \\ \hline
QM & WSB \cite{wsb} &
$0.72$ & $0.68$ & $7.9$ \\
& ISGW \cite{isgw} &
$0.38$ & $0.46$ & -- \\
\hline
\end{tabular}
\caption{\it{Semi-leptonic partial widths (in units of $10^{10}s^{-1}$) for
$D \ra K$, $K^*$, $\pi$, $\rho$ and $\phi$, using $V_{cs}=0.975$ and
$V_{cd}=0.222$. The ratio of the longitudinal to
transverse polarization partial widths for $D \ra K^*$ is also given.
The experimental values of $\Gamma(D^0\ra\pi^-)$, and $\Gamma(D_s\ra\phi)$
have been computed by taking the corresponding branching ratios and
meson life-times from ref. [26].}}
\label{tab:larghezze}
\end{table}
Our central values for the four relevant form factors are in  good
agreement with the experimental data (typically to within less than $10\%$),
although we still suffer, in some cases, from sizeable statistical errors,
especially in the case of  $A_2$.  Considering that  lattice
calculations do not contain free parameters, we find this agreement
 remarkable.  We also observe (tables
\ref{tab:formfac} and \ref{tab:larghezze}) that  predictions
from  $QCD$ sum rules calculations are in agreement with lattice calculations
and experimental determinations, whereas
quark models fail to describe the $D\rightarrow K^{*}$ decay.
Besides the results reported in the tables, we have also obtained
$f_+^\pi(0)/f_+^K(0) = 1.02 \pm  0.03$.
\subsection{Extrapolation to $B-$Meson Decay}

At the values of lattice spacing currently used in numerical simulations,
we are unable to study directly the b quarks. However, as discussed
above, in order to obtain
indirect information on $B-$meson semi-leptonic decays, we can follow the
strategy suggested in ref. \cite{elc}:
we study the form factors in the region of the charm quark mass and then
extrapolate the results to the bottom mass by using the scaling behaviour
predicted by the HQET, eqs. (\ref{scala}). In order to reduce the
uncertainty due to the extrapolation, one could also
compute the form factors in the static limit, i.e. the limit in which the
heavy quark mass is infinite. This determination is not available yet.

\begin{figure}[t]   
\begin{center} \setlength{\unitlength}{1truecm}
\begin{picture}(6.0,6.0)
\put(-6.0,-6.2){\special{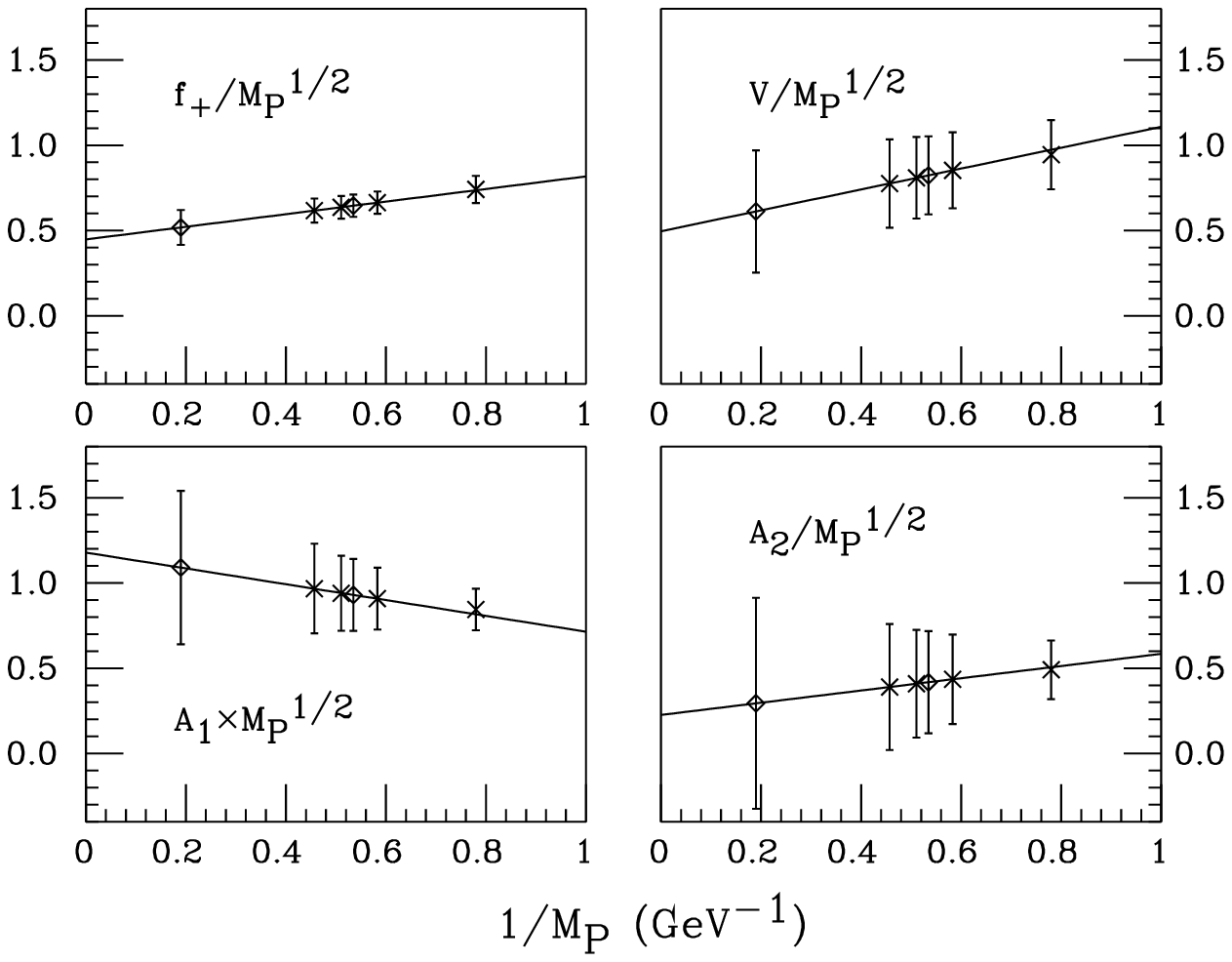}}
\end{picture} \end{center}
\vskip 2.6cm
\caption{\it{Extrapolation of the form factors to the $D$- and $B$-meson
masses (diamonds) using the predictions of the HQET. The points
corresponding to the lightest heavy quark mass have not been used in the
fits.}}
\label{fig:scala}
\end{figure}

Our results show that
the dependence of the form factors on the heavy meson mass, $M_P$, is
compatible with the HQET predictions, see eq.(\ref{scala}), as first
observed in ref. \cite{elc}. This dependence is shown in fig. \ref{fig:scala},
where the four relevant form factors,
extrapolated to the chiral limit in the light quark mass, are given,
 as a function of $1/M_P$ (crosses).
The form factors  were computed  with  $\vec p_D=0$
and  $\vec p_{K,K^*}=2\pi /La (1,0,0)$.
 The values interpolated/extrapolated to the
$D$ and $B$ meson masses (diamonds) are also given.

{}From the values of fig. \ref{fig:scala}, we can compute the form
factors at $q^2=0$. They have been obtained
 by assuming the  pole meson dominance behaviour, using the meson masses
 given in table \ref{tab:masses}.
The corresponding results are presented in table \ref{tab:bdecays}, labelled
``b'', together with the results of other theoretical determinations.
As discussed earlier, in order to evaluate the
stability of these results with respect to a different extrapolation, we
 also give the values (labelled as ``a'')
 obtained with the na\"\i ve scaling laws, eq.
 (\ref{naive}).%
\begin{table}
\centering
\begin{tabular}{|c|c||c|c|c|}
\hline
\multicolumn{2}{|c||}{\bf Reference} & {\bf f$_+$(0)} & {\bf V(0)} & {\bf
A$_1$(0)} \\
\hline \hline
LAT & {\bf This work ``a''} &
$0.29 \pm 0.06$ & $0.45 \pm 0.22$ & $0.29 \pm 0.16$ \\
& {\bf This work ``b''} &
$0.35 \pm 0.08$ & $0.53 \pm 0.31$ & $0.24 \pm 0.12$ \\
& ELC ``a'' \cite{elc} &
$0.26 \pm 0.12\pm 0.04 $ & $0.34 \pm 0.10$ & $0.25 \pm 0.06$ \\
& ELC ``b'' \cite{elc} &
$0.30 \pm 0.14\pm 0.05 $ & $0.37 \pm 0.11$ & $0.22 \pm 0.05$ \\ \hline
SR & Ball \cite{ball} &
$0.26 \pm 0.02$ & $0.6 \pm 0.2$ & $0.5 \pm 0.1$ \\ \hline
QM & WSB \cite{wsb} &
$0.33$ & $0.33$ & $0.28$ \\
& ISGW \cite{isgw} &
$0.09$ & $0.27$ & $0.05$ \\ \hline \hline
\multicolumn{2}{|c||}{\bf Reference} & {\bf A$_2$(0)} & {\bf
V(0)/A$_1$(0)} & {\bf A$_2$(0)/A$_1$(0)} \\
\hline \hline
LAT & {\bf This work ``a''} &
$0.24 \pm 0.56$ & $2.0 \pm 0.9$ & $0.8 \pm 1.5$ \\
& {\bf This work ``b''} &
$0.27 \pm 0.80$ & $2.6 \pm 1.9$ & $1.0 \pm 3.1$ \\
& ELC ``a'' \cite{elc} &
$0.38 \pm 0.18 \pm 0.04$ & $1.4 \pm 0.2$ & $1.5 \pm 0.5 \pm 0.2$ \\
& ELC ``b'' \cite{elc} &
$0.49 \pm 0.21 \pm 0.05$ & $1.6 \pm 0.3$ & $2.3 \pm 0.7 \pm 0.2$ \\
\hline
SR & Ball \cite{ball} &
$0.4 \pm 0.2$ & -- & -- \\ \hline
QM & WSB \cite{wsb} &
$0.28$ & $1.2$ & $1.0$ \\
& ISGW \cite{isgw} &
$0.02$ & $5.4$ & $0.4$ \\ \hline
\end{tabular}
\caption{\it{Semi-leptonic form factors for $B \rightarrow \pi$ and
$\rho$. The label ``a" refers to the na\"\i ve extrapolation in $1/M_P$,
eq. (3), and label ``b" to the extrapolation given in eq. (4).
To extrapolate to zero momentum transfer we have used the masses
of table 1. We have taken the ISGW form factors, as extrapolated to $q^2=0$,
in ref. [17].}}
\label{tab:bdecays}
\end{table}
In the case of the $D-$meson the difference between the HQET scaling laws
and na\"\i ve scaling is immaterial. For the $B-$meson, the differences are
smaller than the statistical errors (typically $10-20\%$), so that we are
not able to distinguish between the two behaviours. We find reassuring,
however, that the results for $B-$decays are in good agreement with the
results of ref. \cite{elc}, where the standard fermion Wilson action and a
smaller lattice spacing, corresponding to $\beta =6.4$, were used. Finally, we
observe that we find values for $A_1$ and $A_2$ smaller than those
obtained by  $QCD$
sum rules, cf. table \ref{tab:bdecays}. The reason is probably due to the
fact that we have assumed pole dominance, whereas $QCD$ sum rules find that the
axial form factors are flat in $q^2$.

{}From the values of the form factor $f_+(0)$  given in table
\ref{tab:bdecays}, and by assuming meson dominance for the $q^2$-dependence,
 we can give an estimate of the $B \rightarrow \pi$ decay rate. To
evaluate the errors, we have allowed the form factor to vary in all
possible ways by one $\sigma$ within the statistical errors and to vary
in all possible ways among the values obtained with different
extrapolations in $1/M_P$, i.e. fits ``a" and ``b". We obtain
\begin{equation}
\Gamma(B \rightarrow \pi l \nu_l)= \vert V_{ub} \vert^2 \,
(8 \pm 4) \times 10^{12} s^{-1},
\end{equation}
that corresponds to the branching ratio
\begin{equation}
B(B \rightarrow \pi l \nu_l)= \vert V_{ub} \vert^2 \, (12 \pm 6)
\end{equation}
when the value $\tau _B = (1.49 \pm 0.12) \times 10^{-12}$ sec. is used
for the $B-$meson lifetime.
On the other hand, with our present accuracy, the errors on the form
factors for the $B \rightarrow \rho$ decay are still too large to get an
estimate of the corresponding branching ratio. A accurate
determination of this quantity can  only be achieved by working with heavier
quark masses and by going to larger lattices.

\section*{Acknowledgements}
\par
We thank R. Gupta, L. Lellouch, J. Nieves and C.T. Sachrajda for discussions.
We acknowledge the partial support by  M.U.R.S.T. and by the EC contract
CHRX-CT92-0051. C.R.A. acknowledges the support by the EC Human Capital and
Mobility Program, contract ERBCHBICT941462.

\end{document}